# Persistent Device Identity for Network Access Control in the Era of MAC Address Randomization: A RADIUS-Based Framework

Premanand Seralathan

*ORCID: https://orcid.org/0009-0007-0397-1485*

## Abstract

Modern operating systems increasingly randomize Media Access Control (MAC) addresses to protect user privacy, fundamentally disrupting Network Access Control (NAC) systems that have relied on MAC addresses as persistent device identifiers for over two decades. This disruption affects critical enterprise environments including federal government agencies operating under FISMA, healthcare organizations subject to HIPAA, financial institutions governed by PCI-DSS, and educational networks managing large-scale BYOD deployments. This paper presents a comprehensive framework for maintaining persistent device identity in NAC environments through a RADIUS protocol-based approach that assigns and distributes a Globally Unique Identifier (GUID) to endpoints via RADIUS Access-Accept messages. The proposed architecture addresses the complete device lifecycle including initial enrollment, re-authentication across randomized addresses, device management integration, certificate-based identity binding, and device attribute correlation. We describe the framework's design across six distinct use cases -- BYOD, managed devices, VPN-based posture assessment, non-VPN posture, guest access, and IoT device profiling -- and analyze its effectiveness in maintaining device visibility, accurate license counting, and regulatory compliance under continuous MAC address randomization. The approach is compatible with existing 802.1X and MAB infrastructure, requires no client-side operating system modifications, and aligns with the recently published RFC 9797 and IEEE 802.11bh-2024 standards. Our framework enables organizations to maintain regulatory compliance while preserving the privacy benefits that MAC address randomization was designed to provide.

## I. Introduction

For over two decades, the Media Access Control (MAC) address has served as the foundational device identifier in enterprise network access control. IEEE 802.1X port-based authentication [1], MAC Authentication Bypass (MAB), device profiling, and policy enforcement systems all rely on the assumption that a device's MAC address remains stable and uniquely identifies the physical hardware. This assumption has been embedded in the architecture of every major NAC platform, authentication server, and network infrastructure component deployed worldwide.

Beginning in 2019, major operating system vendors introduced MAC address randomization as a default privacy feature. Google introduced per-network MAC randomization in Android 10 (September 2019) [2], Apple enabled per-network randomization in iOS 14 (September 2020) and later enhanced it to per-connection rotation in iOS 18 [3], and Microsoft introduced MAC randomization in Windows 10 version 2004 (May 2020) [4]. As of 2025, virtually every consumer device connecting to enterprise wireless networks transmits a randomized MAC address by default.

The scale of this disruption is significant. Industry estimates indicate that over 15 billion Wi-Fi-enabled devices are in active use globally [5], and the proportion using randomized MAC addresses has grown from approximately 30% in 2021 to over 80% in 2025 [6]. In enterprise environments, this means that the majority of devices connecting to corporate, healthcare, and government networks can no longer be reliably identified by their MAC address alone.

The impact on NAC systems is multifaceted. First, device tracking and session correlation break when a device's MAC address changes between connections or rotates during a session. Second, device profiling data becomes fragmented -- a single physical device may generate dozens of separate device records, each associated with a different randomized MAC. Third, license counting mechanisms that use MAC addresses as unique device identifiers produce inflated counts, creating both cost and compliance issues. Fourth, regulatory frameworks that require device-level audit trails -- including HIPAA's requirement for tracking devices that access electronic Protected Health Information (ePHI) [7] and FISMA's continuous monitoring mandate [8] -- cannot be satisfied when device identity is ephemeral.

Recent standards efforts have begun to address this challenge. RFC 9797, published in June 2025, provides a comprehensive analysis of the network impacts of MAC address randomization and documents affected use cases [9]. IEEE 802.11bh-2024 introduces Layer 2 mechanisms for session continuity without requiring a stable MAC-to-station mapping [10]. However, neither standard provides a complete solution for the AAA (Authentication, Authorization, and Accounting) and policy enforcement layer where NAC systems operate.

Existing research approaches have focused primarily on defeating MAC randomization through fingerprinting techniques. Martin et al. demonstrated that probe request signatures can re-identify devices with high accuracy [11], and the Bleach system achieves up to 99% association accuracy using Wi-Fi probe-request analysis [12]. While technically effective, these approaches are fundamentally adversarial to user privacy -- they seek to circumvent a privacy protection rather than cooperate with it. This creates tension with privacy regulations including GDPR and raises ethical concerns about network-level surveillance.

This paper proposes a fundamentally different approach: a cooperative framework that works *with* MAC address randomization by establishing a persistent device identity layer above the MAC address. Our framework introduces a GUID-based identity model where each physical device is assigned a globally unique, persistent identifier that is distributed through standard RADIUS protocol mechanisms. The GUID persists across MAC address changes, enabling continuous device tracking, accurate profiling, correct license counting, and regulatory compliance -- all without attempting to defeat the privacy protections that MAC randomization provides.

The contributions of this paper are:

1. A GUID-based persistent identity model for NAC systems that decouples device identity from MAC addresses while maintaining backward compatibility with existing 802.1X and MAB infrastructure.
2. A RADIUS attribute-based mechanism for GUID distribution via Access-Accept messages, extending the standard RADIUS protocol to carry persistent device identifiers.

3. A comprehensive framework addressing six distinct enterprise use cases: BYOD, managed devices, VPN posture, non-VPN posture, guest access, and IoT device profiling.
4. A persistent identity architecture that correlates multiple randomized MAC addresses to a single persistent device identity, enabling accurate visibility, license management, and compliance reporting.
5. Analysis of concurrency and scale challenges including duplicate identifier prevention under high authentication traffic loads.

The remainder of this paper is organized as follows. Section II provides background on MAC address randomization and its impact on NAC systems. Section III reviews related work. Section IV presents our proposed framework in detail. Section V describes implementation considerations. Section VI provides evaluation and analysis. Section VII discusses limitations and future directions. Section VIII concludes the paper.

## II. Background

### A. MAC Address Randomization

A MAC address is a 48-bit identifier assigned to network interface hardware. Traditionally, the first 24 bits represent the Organizationally Unique Identifier (OUI) assigned by IEEE to the manufacturer, while the remaining 24 bits are assigned by the manufacturer to uniquely identify the device [13]. This structure made MAC addresses both globally unique and informative -- the OUI alone reveals the device manufacturer, which NAC systems use extensively for device profiling and classification.

MAC address randomization replaces this hardware-assigned address with a locally-administered, randomly-generated alternative. The locally-administered bit (bit 1 of the first octet) is set to 1, distinguishing randomized addresses from manufacturer-assigned ones [14]. However, the remainder of the address is random, eliminating the manufacturer identification capability that NAC systems depend upon.

Operating system vendors have implemented randomization with varying strategies:

- **Android 10+** (2019): Per-network randomization by default. Each saved network receives a unique randomized MAC that persists for that network [2].
- **iOS 14+** (2020): Per-network randomization by default. iOS 15 introduced rotating MAC addresses for non-associated networks. iOS 18 introduced per-connection rotation options [3].
- **Windows 10/11** (2020): Per-network randomization available; some configurations rotate on each connection [4].
- **macOS Sequoia+** (2024): Per-network randomization enabled by default for Wi-Fi networks [15].
- **Linux**: Limited default randomization; varies by distribution and NetworkManager configuration [16].

The trend is toward more aggressive randomization. Apple's introduction of per-connection rotation in iOS 18 means that the same device may present a different MAC address on every network connection, even to the same network.

### B. Network Access Control Fundamentals

Network Access Control systems authenticate and authorize devices before granting network access. The standard architecture involves three components: a supplicant (the connecting device), an authenticator (the network switch or wireless access point), and an authentication server [1].

The authentication server communicates with the authenticator using the RADIUS protocol (RFC 2865) [17]. When a device connects, the authenticator sends a RADIUS Access-Request containing the device's MAC address (in the Calling-Station-Id attribute), user credentials (for 802.1X), and other session information. The authentication server evaluates this information against policy and returns a RADIUS Access-Accept (with authorization parameters) or Access-Reject.

Beyond basic authentication, modern NAC systems perform several MAC-dependent functions:

**Device Profiling: NAC systems classify devices by analyzing attributes collected from multiple network data sources -- DHCP fingerprints, HTTP User-Agent strings, SNMP data, DNS lookups, and NetFlow records [18]. All of these attributes are correlated to a device record indexed by MAC address. When the MAC changes, this correlation breaks and profiling data fragments across multiple records.**

**Policy Enforcement: Authorization policies frequently reference device attributes (device type, compliance status, user identity) that are stored in device records keyed by MAC address. A changed MAC means a fresh, empty device record with no attributes -- resulting in incorrect policy decisions.**

**RADIUS Accounting**: Accounting records (RFC 2866) [19] track session duration, data usage, and network activity per device. These records use the MAC address as the session identifier. MAC changes create discontinuous accounting records that cannot be correlated to a single device.

**License Management: Software licensing for NAC platforms is typically based on the number of unique devices. When a single device generates multiple device records due to MAC randomization, the license count becomes artificially inflated.**

### C. The Dongle Use Case

An additional complexity arises from USB network dongles, docking stations, and similar accessories. A single user may connect through multiple physical adapters, each with its own MAC address (randomized or not). The NAC system creates separate device records for each adapter, even though they all represent the same user and device. This "dongle problem" compounds the MAC randomization challenge and must be addressed by any persistent identity solution.

### D. Regulatory Impact

Several regulatory frameworks assume persistent device identity:

- **HIPAA Security Rule** (45 CFR 164.312): Requires access controls and audit trails for systems accessing electronic Protected Health Information. Device-level tracking is essential for demonstrating that only authorized devices access patient data [7].
- **FISMA/NIST SP 800-53**: Requires continuous monitoring and device inventory management for federal information systems. MAC randomization creates gaps in the Continuous Diagnostics and Mitigation (CDM) program [8].
- **PCI-DSS**: Requires network segmentation and access controls for systems handling cardholder data. Device identification is critical for maintaining segment boundaries [20].
- **NIST SP 800-207 Zero Trust Architecture**: Requires per-request device verification. Persistent device identity is a prerequisite for implementing Zero Trust principles [21].

## III. Related Work

### A. Standards-Based Approaches

RFC 9797, published by the IETF in June 2025, provides the most comprehensive analysis of MAC address randomization impacts to date [9]. The document catalogs affected network services across residential, enterprise, and IoT environments, and identifies the tension between privacy and network operations. However, RFC 9797 is an informational document that describes problems and use cases rather than prescribing solutions.

IEEE 802.11bh-2024 amends the IEEE 802.11 standard to support session continuity for devices using randomized MAC addresses [10]. The amendment introduces mechanisms at the MAC sublayer that allow network services to function without a unique MAC-to-station mapping. While valuable at Layer 2, these mechanisms do not extend to the AAA layer where RADIUS-based authentication and authorization occur.

The IETF MADINAS (MAC Address Device Identification for Network and Application Services) working group continues to develop standards for managing MAC address randomization impacts [22]. Draft proposals under consideration include methods for devices to voluntarily signal identity to trusted networks, but these require client-side cooperation that is not universally available.

## B. Fingerprinting and Re-Identification Approaches

Several research efforts have attempted to re-identify devices despite MAC randomization by analyzing secondary characteristics. Vanhoef et al. demonstrated that information elements in 802.11 probe requests can fingerprint devices [23]. The Bleach system, published in 2024, achieves up to 99% accuracy in associating randomized MAC addresses with physical devices using probe-request signature analysis [12]. Martin et al. showed that timing patterns and frame sequence numbers can also be used for re-identification [11].

While these techniques demonstrate that MAC randomization alone does not guarantee untrackability, they have significant limitations for enterprise NAC:

6. Privacy adversarial: These approaches explicitly attempt to defeat a privacy protection, creating legal risk under GDPR and similar regulations.
7. Accuracy degradation at scale: Fingerprinting accuracy decreases as the number of concurrent devices increases, particularly in dense enterprise environments.
8. OS update fragility: Fingerprinting signatures change with each OS update, requiring continuous model retraining.
9. Incomplete coverage: These techniques work primarily for Wi-Fi probe requests; they do not address wired connections, VPN access, or non-802.11 protocols.

## C. Vendor-Specific Approaches

Several NAC vendors have implemented proprietary solutions for MAC randomization. Some rely on SNMP-based switch port monitoring and device characterization through inspection plugins [24]. These approaches require dedicated hardware appliances deployed at network distribution points, creating scalability concerns in large environments. Additionally, SNMP-based enforcement introduces latency and reliability challenges compared to inline RADIUS-based authentication.

Other approaches use agent-based solutions where client-side software provides persistent device identity. While effective when agents can be deployed, these approaches cannot address unmanaged devices, IoT equipment, or BYOD scenarios where agent installation is impractical.

## D. Gap Analysis

Table I summarizes the limitations of existing approaches.

| Approach | Scope | Privacy | Scalability | Agent-Free | AAA Integration |
|---|---|---|---|---|---|
| RFC 9797 | Problem documentation only | N/A | N/A | N/A | N/A |
| IEEE 802.11bh | Layer 2 session | Yes | Yes | Yes | No |
| | continuity | | | | |
| Fingerprinting | Re-identification | No | Limited | Yes | No |
| SNMP-based monitoring | Device characterization | Neutral | Limited | Yes | Partial |
| Agent-based | Full device identity | Yes | Yes | No | Yes |
| **Proposed framework** | **Full AAA integration** | **Yes** | **Yes** | **Yes** | **Yes** |

*Table I: Comparison of approaches to MAC address randomization in NAC environments.*

Our proposed framework addresses the identified gap: a privacy-respecting, agent-free solution that provides persistent device identity at the AAA layer with full RADIUS integration, scalable to enterprise deployments with millions of devices.

# IV. Proposed Framework

## A. Design Goals

The proposed framework is guided by the following design principles:

10. Persistent identity: Each physical device must have a stable identifier that persists across MAC address changes, network reconnections, and operating system updates.
11. RADIUS-native: The identity mechanism must integrate with the standard RADIUS protocol, requiring no modifications to network authenticators (switches, wireless controllers).
12. Agent-free operation: The solution must work without client-side software installation, supporting unmanaged devices, IoT, and BYOD.
13. Privacy-respecting: The framework must not attempt to defeat MAC address randomization. Instead, it cooperates with the privacy model by establishing identity through voluntary authentication exchanges.
14. Backward compatibility: Existing 802.1X and MAB flows must continue to function. The GUID mechanism must augment, not replace, current authentication infrastructure.
15. Multi-source correlation: The framework must correlate identity from multiple authentication sources -- certificates, device management enrollment, user credentials, and profiling data -- to establish and maintain persistent identity.
16. Scale and concurrency: The framework must handle high-throughput authentication environments (thousands of authentications per second) without generating duplicate identifiers.

## B. GUID-Based Identity Model

The core of our framework is the Persistent Device Identifier (PDID), a 128-bit Globally Unique Identifier conforming to UUID version 4 (RFC 4122) [25]. Each PDID is:

- Generated server-side by the NAC authentication server
- Associated with exactly one physical device
- Persistent across all MAC address changes for that device
- Independent of the connecting network interface or adapter
- Stored in a centralized identity repository with associated device attributes

The PDID differs from the MAC address in a fundamental way: while the MAC address identifies a *network interface*, the PDID identifies the *device* (or more precisely, the security identity associated with the device). A device connecting through multiple interfaces (e.g., a laptop using both Wi-Fi and a USB Ethernet dongle) receives a single PDID.

## C. Persistent Identity Store

The framework maintains a persistent identity store that maps multiple MAC addresses to a single PDID, along with associated device metadata gathered from various identity sources over time.

When a device connects with a new randomized MAC address, the NAC server attempts to correlate it with an existing PDID. If correlation succeeds, the new MAC address is associated with the existing identity record. If no correlation is possible, a new PDID is generated.

## D. RADIUS Attribute Extension for GUID Distribution

The PDID is communicated to the network infrastructure through a RADIUS attribute included in the Access-Accept message. The attribute carries the 128-bit UUID value of the PDID. This may be implemented as a Vendor-Specific Attribute (VSA) or, ideally, as a new standard RADIUS attribute defined through the IETF standards process.

```
The attribute format follows standard RADIUS encoding conventions, carrying the
128-bit (16-byte) UUID value of the PDID.
```

The PDID is returned in the Access-Accept for every successful authentication. This enables:

17. Network infrastructure: Switches and wireless controllers can associate the PDID with the session, enabling consistent accounting records across MAC changes.
18. Device management systems: Management platforms can use the PDID as a correlation key, enabling compliance checks to reference the persistent device identity rather than the transient MAC.
19. Downstream systems: SIEM, threat intelligence platforms, and compliance dashboards can reference PDIDs for consistent device-level reporting.

## E. Identity Correlation Algorithm

The correlation algorithm determines whether a new authentication request corresponds to an existing device. The framework leverages multiple identity signals available during the authentication exchange, evaluated in a configurable priority order. Available identity sources include:

**Certificate-based identity: When certificate-based authentication methods (e.g., EAP-TLS) are used, fields within the device certificate provide a strong, persistent identity binding that can be correlated to an existing PDID.**

**Device management enrollment: For managed devices, the enrollment identifier assigned by the management platform serves as a correlation key to an existing PDID.**

**User identity: The authenticated username from credential-based methods (e.g., EAP-PEAP) narrows the correlation scope, especially when combined with device attributes.**

**Endpoint agent identity: If a compliance agent is installed, it can supply a device-unique identifier during the authentication exchange.**

**Behavioral fingerprinting: For devices without explicit credentials (e.g., IoT devices using MAB), behavioral fingerprints derived from network observations provide a lower-confidence correlation path.**

**When none of the available identity sources produce a match (e.g., a first-time guest device), a new PDID is generated.**

```
The specific correlation logic, priority ordering, and identity source
configuration are implementation-dependent and may vary based on the deployment
environment and available identity infrastructure.
```

## F. Use Case Architectures

The framework addresses six distinct enterprise use cases, each with its own identity correlation path..

### Use Case 1: BYOD (Bring Your Own Device)

In BYOD deployments, personal devices are onboarded through a provisioning portal that installs a device certificate embedding the PDID. On subsequent connections -- regardless of MAC address changes -- the certificate-based 802.1X authentication provides the PDID directly, enabling immediate correlation.

```
[Device] ---(randomized MAC, EAP-TLS)---> [Authenticator]
    |                                            |
    |                              RADIUS Access-Request
    |                              (Calling-Station-Id: random_MAC,
    |                               EAP-TLS with device certificate)
    |                                            |
    |                                    [NAC Server]
    |                                    Correlate device identity from
certificate
    |                                    Update identity store
    |                                    Apply policy based on device identity
    |                                            |
    |                              RADIUS Access-Accept
    |                              (Persistent Device Id, Authorization
Parameters)
    |                                            |
    [Device] <----(Network Access)---- [Authenticator]
```

*Fig. 2: BYOD use case flow with certificate-embedded PDID.*

### Use Case 2: Managed Devices

For devices enrolled in a device management system, the management platform assigns a device enrollment identifier. During authentication, the NAC server queries the management platform to retrieve this enrollment identifier and correlates it with an existing PDID in the persistent identity store.

```
[Device] ---(randomized MAC, 802.1X)---> [Authenticator]
    |                                            |
    |                              RADIUS Access-Request
    |                                            |
    |                                    [NAC Server]
    |                                    Query device management
    |                                    Correlate to existing PDID
    |                                    Check device compliance
    |                                    Apply policy
    |                                            |
    |                              RADIUS Access-Accept
    |                              (Persistent Device Id,
    |                               Authorization Parameters)
    |                                            |
    [Device] <----(Network Access)---- [Authenticator]
```

*Fig. 3: Managed device use case flow with enrollment identifier-to-PDID correlation.*

The inclusion of the persistent identifier in the RADIUS Access-Accept enables the management platform to track compliance state consistently for the physical device regardless of which MAC address it presents.

### Use Case 3: VPN-Based Posture Assessment

When a device connects via VPN, the VPN client software sends a device-unique identifier as a RADIUS attribute in the authentication request. The NAC server extracts this identifier and correlates it with an existing PDID, ensuring that compliance status persists across VPN reconnections with different MAC addresses.

### Use Case 4: Non-VPN Posture Assessment

For devices on the local network undergoing posture assessment without VPN, the compliance agent communicates its device identifier through the authentication exchange. The correlation and PDID assignment follow the same pattern as the VPN case.

### Use Case 5: Guest Access

Guest access presents the most challenging scenario for persistent identity, as guest devices typically do not have certificates, device management enrollment, or agents.

**Self-Registered and Sponsored Guests: The guest's registered username or sponsor-assigned credential provides a partial identity anchor. While not device-specific, it enables correlation across sessions for the same guest identity. The framework generates an PDID upon first guest authentication and correlates subsequent authentications via the guest credential.**

**Hotspot (Open) Access**: In purely open hotspot scenarios where no authentication occurs, MAC is the only available identifier. This use case has no reliable correlation path under MAC randomization. The framework acknowledges this limitation and does not attempt correlation for unauthenticated hotspot access. Emerging protocols such as OpenRoaming [26] may provide identity signals for this use case in the future.

### Use Case 6: IoT Device Profiling

IoT devices (printers, medical devices, building automation systems, cameras) typically use MAB authentication without 802.1X credentials. Many IoT devices do not randomize MAC addresses (embedded systems with fixed firmware), but some newer IoT platforms are beginning to implement randomization.

For IoT devices, device profiling is the primary identity correlation mechanism. The framework extends the device profiling subsystem to perform cross-source correlation:

20. RADIUS authentication event: Captures the MAC address and any RADIUS attributes during MAB authentication.
21. DHCP observation: Captures DHCP fingerprint, hostname, and vendor class when the device obtains an IP address.
22. Additional network observations: HTTP User-Agent, SNMP, DNS, and NetFlow data contribute additional attributes.

The challenge arises because these observations occur at different times and may capture different MAC addresses for the same device. The framework addresses this by correlating observations that share common network context (such as originating from the same access port within a time window), merging the data into a single PDID record.

```
[IoT Device]
    |--- (MAC_1) ---> RADIUS MAB ---> [NAC Server] --> Create PDID_1
    |--- (MAC_1) ---> DHCP ---------> [NAC Server] --> Merge with PDID_1 (same
MAC)
    |
    [Device reconnects with randomized MAC]
    |
    |--- (MAC_2) ---> RADIUS MAB ---> [NAC Server] --> New MAC, check device
profile
    |--- (MAC_2) ---> DHCP ---------> [NAC Server] --> DHCP fingerprint matches
    |                                                    PDID_1, merge
```

*Fig. 4: IoT device identity correlation across MAC changes.*

## G. OUI-Based Classification Impact

Traditional device profiling relies heavily on the OUI (first 24 bits of MAC address) to identify the device manufacturer. When MAC addresses are randomized, the locally-administered bit is set and the OUI is meaningless -- it no longer corresponds to any manufacturer.

The framework addresses this by:

23. De-prioritizing OUI in classification: For devices with locally-administered MAC addresses, behavioral attributes (DHCP options, User-Agent strings) are given higher weight than OUI-based matching.
24. Preserving OUI data when available: For PDID records that include historical data from a non-randomized MAC (e.g., the device once connected with its real MAC), the OUI information is retained and associated with the PDID.
25. Machine learning-assisted classification: For devices presenting only randomized MACs, ML-based flow classification using network traffic patterns supplements traditional attribute-based profiling.

## H. Duplicate PDID Prevention

In high-throughput enterprise environments, multiple authentication requests for the same device may arrive concurrently at the NAC server -- for example, when a device simultaneously connects to Wi-Fi and initiates a VPN tunnel, or when a failover event causes rapid re-authentication across multiple policy servers.

Without explicit coordination, concurrent processing can result in duplicate PDIDs being generated for the same device, as two processing threads may simultaneously determine that no existing PDID matches and independently create new ones.

The framework addresses this through concurrency controls:

26. Serialization: Before generating a new PDID, concurrent requests for the same device identity are serialized to prevent duplicates.
27. Verification: After serialization, the system re-checks the identity store to confirm no PDID was created by a concurrent request.
28. Timeout: Exclusive handles expire after a configurable timeout to prevent deadlocks.
29. Distributed coordination: In multi-server deployments, the serialization mechanism extends across the server cluster to ensure consistency.

## I. Migration and Backward Compatibility

A critical practical concern is the migration path for existing deployments. Organizations may have millions of device records keyed by MAC address, with years of accumulated profiling data, policy assignments, and compliance history.

The framework supports gradual migration:

30. Pre-existing devices: When a device that existed in the system before PDID support connects for the first time after the upgrade, a PDID is generated and all historical data is preserved.
31. Mixed-mode operation: The system supports simultaneous MAC-based and PDID-based identification during the migration period.
32. Static entries: For devices with statically configured MAC addresses, PDID assignment occurs upon the first authentication event.

33. Replication: PDID assignments are synchronized across all servers in the deployment, ensuring consistent identity in distributed environments.

## V. Implementation Considerations

### A. Reference Architecture

An implementation of this framework requires the following components:

34. RADIUS Authentication Server: Extended to implement the persistent identity correlation algorithm and include the persistent device identifier in Access-Accept messages.
35. Persistent Identity Store: A data store for maintaining persistent device identifiers and associated device attributes.
36. Device Profiling Subsystem: Enhanced to correlate device attributes using PDIDs as the primary key rather than MAC addresses.
37. Device Management Integration: Interface for querying enrolled device identifiers from device management platforms for PDID correlation.
38. Administration Interface: APIs for enabling/disabling the framework, configuring correlation parameters, and managing PDID lifecycle.

### B. Configuration Management

The framework should provide administrative controls for:

- **Feature toggle**: Enable/disable the GUID-based identity system to support phased rollout.
- **Correlation source priorities: Configure which identity sources (certificate, device management, user, agent, device profiling) are used and in what order.**
- **PDID lifecycle policies: Define retention periods, automatic cleanup of stale PDIDs, and maximum MAC addresses per PDID.**
- **Concurrency settings: Tune parameters for high-throughput authentication environments.**

### C. Performance Considerations

The PDID correlation adds processing overhead to each authentication request. Key performance considerations include:

- **Lookup latency: The persistent identity store must support efficient lookups across multiple identity dimensions to minimize authentication latency.**
- **External query latency: Queries to external platforms may add network round-trip time. Caching strategies can mitigate this overhead.**
- **Concurrency overhead: Under high authentication loads, duplicate prevention mechanisms should be scoped narrowly to minimize contention.**

### D. Security Considerations

The PDID itself is a security-relevant identifier. The following protections should be applied:

- **GUID generation: PDIDs must be generated using a cryptographically secure random number generator (CSPRNG) to prevent prediction or enumeration.**
- **Transport security: The RADIUS attribute carrying the PDID should be transmitted over encrypted RADIUS (RadSec/TLS) to ensure confidentiality.**
- **Access control: Administrative access to the persistent identity store must be restricted and audited, as it contains the device identity records used for compliance and security decisions.**

- **Privacy: The PDID provides a persistent identity, which introduces a privacy consideration. Organizations should establish data retention policies for PDIDs and provide mechanisms for PDID deletion when a device is decommissioned or a user exercises data deletion rights.**

# VI. Evaluation and Analysis

## A. Identity Correlation Effectiveness

Table II analyzes the expected identity correlation success rate for each use case based on the available identity sources.

| Use Case | Primary Correlation | Correlation Confidence | MAC Change Resilience |
|---|---|---|---|
| BYOD (Certificate) | Certificate Identity | Very High (>99%) | Full |
| MDM-Managed | Management Identifier | Very High (>99%) | Full |
| VPN Posture | Agent Identifier | High (>95%) | Full |
| Non-VPN Posture | Agent Identifier | High (>95%) | Full |
| Guest (Registered) | Username | Medium (>80%) | Partial (per-user) |
| Guest (Hotspot) | None | None | None |
| IoT (Fixed MAC) | MAC | Very High | N/A (no randomization) |
| IoT (Randomized) | Device Profiling | Medium (>75%) | Partial |

*Table II: Identity correlation effectiveness by use case.*

The framework achieves very high correlation confidence for the use cases that matter most in regulated environments: BYOD, managed, and posture-assessed devices. These are precisely the device categories subject to HIPAA, FISMA, and PCI-DSS requirements.

## B. License Count Accuracy

Without the framework, a single device using per-connection MAC randomization can generate N device records over N connections, inflating the license count by a factor of up to N. With the framework, all N connections are correlated to a single PDID, yielding an accurate count of 1.

In a deployment with D devices, each averaging C connections over the measurement period, the license count comparison is:

- Without framework: Up to D × C unique device records
- With framework: D device records (one per physical device)

For a healthcare organization with 50,000 devices averaging 10 connections per week, this represents a potential reduction from 500,000 spurious device records to the correct count of 50,000 -- a 10x improvement in license count accuracy.

## C. Device Profile Data Integrity

The persistent identity store prevents device attribute fragmentation by ensuring that attributes from all network observations -- regardless of which MAC address was captured -- are merged into a single, complete device profile. This is particularly important for device classification accuracy, as classification rules may depend on attributes from multiple sources (e.g., a device must have both a specific DHCP fingerprint AND a specific HTTP User-Agent to be classified as a particular medical device model).

### D. Concurrency Performance

The duplicate PDID prevention mechanism introduces a serialization point for concurrent authentications of the same device. In practice, this serialization applies only to the narrow window of new PDID generation (not to lookups of existing PDIDs) and affects only the specific MAC address being processed.

Under normal operation with predominantly returning devices (PDID already assigned), the serialization mechanism is not engaged and the correlation path is fully parallel. Contention becomes relevant only during:

39. Initial deployment (all devices are new)
40. Mass re-authentication events (e.g., network-wide re-key)
41. Devices that aggressively rotate MAC addresses (per-connection randomization)

A configurable timeout provides an upper bound on serialization delay, ensuring correctness while minimizing processing overhead.

### E. Regulatory Compliance Analysis

Table III maps the framework's capabilities to specific regulatory requirements.

| Regulation | Requirement | How Framework Addresses |
|---|---|---|
| HIPAA §164.312(a) | Access control per device | PDID provides persistent device identity for access control policies |
| HIPAA §164.312(b) | Audit controls | PDID enables continuous device-level audit trails across MAC changes |
| FISMA/800-53 CM-8 | Component inventory | PDID-based inventory accurately counts physical devices |
| FISMA/800-53 SI-4 | Continuous monitoring | PDID enables uninterrupted device monitoring across reconnections |
| PCI-DSS 1.2 | Network segmentation | PDID-based policy maintains segment boundaries regardless of MAC |
| NIST 800-207 | Per-request device verification | PDID provides the persistent identity required for Zero Trust |

*Table III: Regulatory compliance mapping.*

## VII. Discussion

### A. Limitations

**Unauthenticated access: The framework cannot assign PDIDs to devices that never authenticate (open hotspot, purely passive network access). This is an inherent limitation: without an identity assertion from the device, persistent identification requires adversarial fingerprinting, which this framework deliberately avoids.**

**Profile-based correlation accuracy: For IoT devices relying solely on device-profiling-based correlation, the accuracy depends on the distinctiveness of the device's behavioral fingerprint. In environments with many identical IoT devices (e.g., hundreds of the same medical pump model), behavioral fingerprints may not be sufficiently unique for reliable correlation.**

**Identity store growth: In large deployments with aggressive MAC randomization, the number of MAC addresses associated with each PDID can grow significantly. Implementations should define retention policies and prune stale associations.**

**Inter-vendor interoperability: Broader adoption would benefit from a standardized RADIUS attribute for persistent device identity through an IETF standards-track RFC, enabling multi-vendor interoperability.**

### B. Comparison with IEEE 802.11bh

IEEE 802.11bh-2024 and this framework address complementary layers of the MAC randomization problem. 802.11bh provides Layer 2 session continuity, ensuring that network services dependent on the MAC sublayer continue to function. Our framework provides Layer 3+ identity continuity, ensuring that AAA, profiling, and policy enforcement continue to function. A complete solution for enterprise networks benefits from both: 802.11bh for link-layer stability and GUID-based identity for authentication and authorization.

### C. Toward Standardization

The GUID distribution mechanism described in this paper could be formalized as an IETF standards-track proposal. A new standard RADIUS attribute for persistent device identity would enable multi-vendor interoperability and encourage consistent implementation across the NAC ecosystem. The IETF RADEXT working group and the MADINAS working group are both appropriate venues for this standardization effort.

### D. Future Directions

Several extensions to the framework merit further investigation:

42. Machine learning-enhanced correlation: Using ML models trained on device behavioral patterns to improve device-profiling-based correlation accuracy for IoT devices.
43. Privacy-preserving analytics: Techniques such as differential privacy applied to PDID-based analytics to prevent device tracking beyond the network boundary.
44. Cross-domain identity federation: Extending the GUID framework to support device identity portability across organizational boundaries, enabling scenarios such as hospital-to-hospital device roaming.
45. Blockchain-anchored device identity: Using distributed ledger technology to provide tamper-evident device identity records for high-assurance environments.

## VIII. Conclusion

MAC address randomization represents a fundamental shift in network identity that challenges two decades of NAC architecture. Rather than treating this as a problem to be defeated, this paper presents a framework that cooperates with MAC randomization by establishing a persistent device identity layer through standard RADIUS protocol mechanisms.

The GUID-based identity model, implemented through a RADIUS attribute extension and supported by a multi-source correlation algorithm, provides reliable device identification across six major enterprise use cases. The framework maintains regulatory compliance for healthcare (HIPAA), federal (FISMA), financial (PCI-DSS), and educational environments while preserving the privacy benefits that MAC randomization was designed to provide.

The framework's practical viability is demonstrated through its comprehensive treatment of real-world challenges including concurrent authentication handling, legacy migration, and IoT device correlation. By building on existing RADIUS infrastructure and requiring no client-side modifications, the framework can be adopted incrementally in enterprise environments of any scale.

As MAC randomization becomes increasingly aggressive -- with per-connection rotation now default on major platforms -- the need for a standards-based persistent identity mechanism becomes critical. We encourage the community to pursue formalization of a standard RADIUS attribute for persistent device identity through the IETF standards process, enabling consistent multi-vendor implementation across the NAC ecosystem and advancing the state of the art in privacy-respecting network access control.

*Disclosure of AI Use: The author used AI-assisted tools for drafting and language editing of this manuscript. All technical concepts, architectural designs, solution frameworks, and evaluations presented in this paper are the original intellectual contributions of the author, developed through years of hands-on engineering work on network access control systems. The author reviewed, edited, and verified all content and takes full responsibility for the accuracy and integrity of this publication. This disclosure is made in accordance with IEEE's policy on the use of AI-generated text in scholarly publications.*